\documentclass[referee]{jfm} 

\usepackage{graphicx}
\usepackage{subfigure}
\usepackage{natbib}
\usepackage{amsmath}
\graphicspath{{fig/}}

\ifCUPmtlplainloaded \else
  \checkfont{eurm10}
  \iffontfound
    \IfFileExists{upmath.sty}
      {\typeout{^^JFound AMS Euler Roman fonts on the system,
                   using the 'upmath' package.^^J}%
       \usepackage{upmath}}
      {\typeout{^^JFound AMS Euler Roman fonts on the system, but you
                   dont seem to have the}%
       \typeout{'upmath' package installed. JFM.cls can take advantage
                 of these fonts,^^Jif you use 'upmath' package.^^J}%
      }
  \else
  \fi
\fi

\ifCUPmtlplainloaded \else
  \checkfont{msam10}
  \iffontfound
    \IfFileExists{amssymb.sty}
      {\typeout{^^JFound AMS Symbol fonts on the system, using the
                'amssymb' package.^^J}%
       \usepackage{amssymb}%
         \let\leq=\leqslant
         \let\geq=\geqslant
      }{}
  \fi
\fi

\ifCUPmtlplainloaded \else
  \IfFileExists{amsbsy.sty}
    {\typeout{^^JFound the 'amsbsy' package on the system, using it.^^J}%
     \usepackage{amsbsy}}
    {}
\fi

\newsavebox{\astrutbox}
\sbox{\astrutbox}{\rule[-5pt]{0pt}{20pt}}

\title[]{The motion, stability and breakup of a stretching liquid bridge with a receding contact line}

\author[B. Qian and K. S. Breuer]%
{B\ls I\ls A\ls N\ns Q\ls I\ls A\ls N\ns \footnote{Current address: Department of chemistry, Frick Laboratory, Princeton University, Princeton, NJ 08544, USA}  \and K\ls E\ls N\ls N\ls E\ls T\ls H\ns S.\ns B\ls R\ls E\ls U\ls
E\ls R}

\affiliation{Division of Engineering, Brown University, Providence, RI 02915, USA}

\date{February 2010, Revised July 2010}

\begin{document}

\maketitle

\begin{abstract}
The complex behavior of drop deposition on a hydrophobic surface is considered by looking at a model problem in which the evolution of a constant-volume liquid bridge is studied as the bridge is stretched.  The bridge is pinned with a fixed diameter at the upper contact point, but the contact line at the lower attachment point is free to move on a smooth substrate.  Experiments indicate that initially, as the bridge is stretched, the lower contact line slowly retreats inwards.  However at a critical radius, the bridge becomes unstable, and the contact line accelerates dramatically, moving inwards very quickly.  The bridge subsequently pinches off, and a small droplet is left on the substrate.  A quasi-static analysis, using the Young-Laplace equation, is used to accurately predict the shape of the bridge during the initial bridge evolution, including the initial onset of the slow contact line retraction.  A stability analysis is used to predict the onset of pinch-off, and a one-dimensional dynamical equation, coupled with a Tanner-law for the dynamic contact angle,  is used to model the rapid pinch-off behavior.  Excellent agreement between numerical predictions and experiments is found throughout the bridge evolution, and the importance of the dynamic contact line model is demonstrated.
\end{abstract}

\section{Introduction}
\begin{figure}
\centering
\includegraphics[width=3.5in]{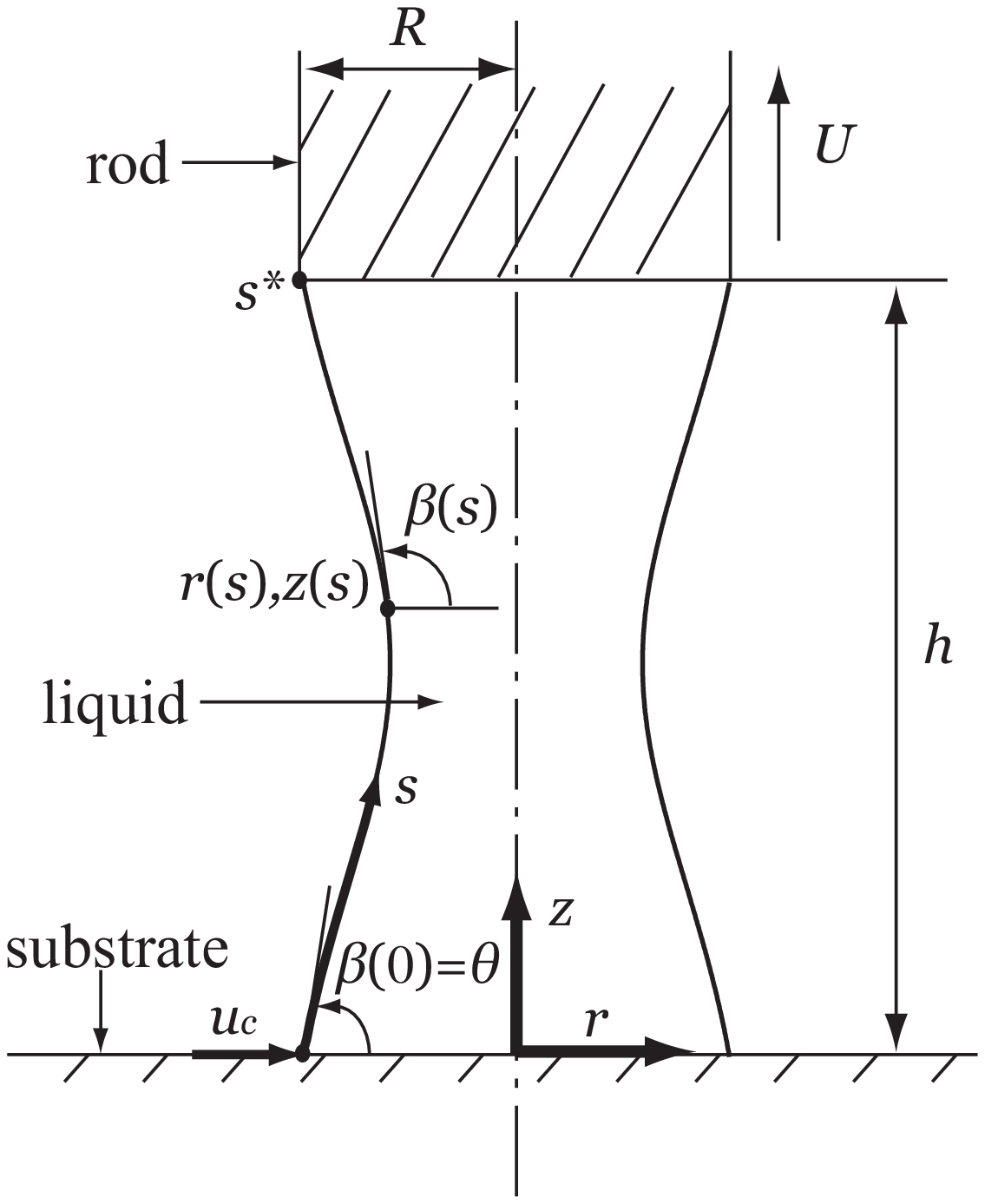}
\caption{Geometry and coordinate system for an axisymmetric liquid bridge with a free moving contact line on the substrate.}
\label{fig:geometry}
\end{figure}

Contact drop dispensing is the process by which a liquid drop may be deposited on a substrate by touching the surface with a wetted tip, such as a needle or a syringe. Although there are a variety of approaches, the basic dispensing process is initiated by bringing a tip close to a flat substrate so that a liquid bridge is formed between the substrate and the dispensing syringe, as sketched in Fig.~\ref{fig:geometry}. As the syringe retreats, the liquid bridge stretches, grows, and breaks, leaving a drop on the substrate.  The technique has many industrial applications, including the printing industry and dispensing of glue for packaging.  Most recently, it has been adapted for a variety of novel uses at small scales, such as direct scanning probe lithography~\cite[]{ginger2003}, micromachined fountain-pen techniques~\cite[]{Deladi2004,moldovan2006} and in the formation of micro-arrays of biological materials~\cite[]{muller2004}. Despite the simplicity of the operation, the exact control of the dispensing drop size is complicated by several factors, such as the syringe geometry, the dispensing speed, the liquid properties and the surface wettability. For the accurate prediction of drop sizes, knowledge of how these factors affect the dispensing process is desirable.  In particular, an accurate prediction of the stability and the breakup of the liquid bridge is needed.

The study of liquid bridges was pioneered over one hundred years ago by Plateau who experimentally investigated the stability of an infinite vertical falling water jet~\cite[]{plateau1863}, in which he observed that the maximum ratio of the stable length to the jet diameter is about a constant $\pi$. The theoretical derivation of the observed stable length limit was given by Rayleigh with a linear stability analysis ~\cite[]{rayleigh1878}, which is known as Rayleigh-Plateau limit. Later broad applications of liquid bridges in industry inspired intensive studies of the stability of a static weightless axisymmetric liquid bridge confined between two circular disks, in which the critical height of the bridge as a function of the bridge volume was theoretically calculated and experimentally tested ~\cite[]{mason1970,gillette1971,sanz1983,Russo1986,slobozanin1993}. The influences of gravity and unequalness of supporting disks on the stable limit were also investigated during the past two decades ~\cite[]{meseguer1985,meseguer1991,slobozanin1993,meseguer1984,meseguer1991,slobozanin1998}. Although the study of static liquid bridges has reached a level of maturity, the investigation of the dynamic stretching of a liquid bridge has had to wait until quite recently,  due to the difficulties in experimentally recording the rapid bridge breakup as well as the complexities associated with the mathematical treatments in theory ~\cite[]{meseguer1985,frankel1985,papageorgiou1995}. The representative work by ~\cite{zhang1996} exhaustively investigated the dependence of breakup features on the stretching speed and the liquid properties with both experimental and numerical methods, and directly compared experiments with theory, finding quantitative agreement, despite the fact that the the calculations were restricted to moderate stretching speeds due to the limitation of the one-dimensional approximate model~\cite[]{eggers1993}. More recently, the numerical calculations have been extended using two-dimensional models and used to investigate, among other things, the effects of different supporting geometries \cite[]{yildirim2001} and the effects of surfactants ~\cite[]{liao2006,Panditaratne_PHD2003} on the breakup of a dynamically stretching liquid bridge. All these previous studies, however,  concentrated on geometries in which both the upper and lower contact lines are pinned.  In contrast, the contact drop dispensing problem is characterized by the fact that the lower contact line is free to move. This modification to the liquid bridge breakup has not been addressed, perhaps due to the lack of accurate models for the dynamic contact line behavior.  However, this frontier too has seen recent progress, including numerical simulations using a diffuse interface model ~\cite[]{amberg2007}, Navier slip boundary condition while maintaining a constant contact angle ~\cite[]{kumar2009}, and an approach using an empirical, velocity-dependent, dynamic contact angle model ~\cite[]{Panditaratne_PHD2003}. These studies have shown that the dynamic contact line is crucial for the breakup dynamics of a stretching liquid bridge and for these reasons, we assume that it has a strong effect on the drop size in the contact dispensing problem.

Previous experiments by our group concerning drop generation on a hydrophobic surface have shown that, by changing the speed of the retracting needle, one can control the contact line motion, and through these means, one can generate a broad range of drop sizes using a single syringe ~\cite[]{qian2009}. The drop dispensing physics can be divided into three regions: advancing contact line, fixed contact line and receding contact line. For low syringe retraction speeds, $U$, the contact line radius slowly expands, and arbitrarily large drops can be generated.  In these regions, the drop sizes,  $r_d$, were shown to vary as $U^{-1/2}$, and due to the low syringe speed and the slow contact line motion, the bridge evolution and breakup was well-predicted by quasi-static theory using Rayleigh-Plateau instability theory, and its extensions for non-cylindrical bridges \cite[]{meseguer1984,slobozhanin1997,slobozanin1998}. However, for higher values of $U$, the contact line recedes, and the process is more complex, and was observed to be comprised of two phases \cite[]{qian2009}: an initial phase characterized by a slow (quasi-static) contact line retraction, followed by a \emph{very} rapid phase in which the contact line speed is comparable to the capillary wave speed, and during which the contact angle is seen to depend on the speed, and to be significantly lower than its quasi-static receding value.

The retreating contact line mode of droplet deposition is of great technical interest, since it allows micron-scale droplets to be deposited using millimeter-scale hardware. However, the complexities of the governing physics are considerable and several questions were left unanswered by the original experiments of \cite{qian2009}.  These questions include determining when the contact line starts to move, at what point does the bridge become unstable and begins to pinchoff, and lastly how the final drop size depends on the liquid-surface interaction.  In this paper, we address these questions by studying in detail the drop dispensing in the receding contact line region using both experimental and numerical tools. To facilitate the study, we have simplified the drop dispensing problem in one aspect, and we consider drops dispensing from a \textit{constant-volume} liquid bridge, instead of a bridge defined by a \textit{constant pressure} at the upper boundary (fed by the flow from a reservoir through a syringe). In our experiment, the dispensing drop sizes and the contact line motions were measured for different dispensing speeds and liquid volumes. Our theoretical analysis is divided into three components: (a) a quasi-static analysis using the Young-Laplace equation to describe the initial bridge evolution (although, still allowing for slow contact line motion); (b) a stability analysis to predict the onset of the bridge pinchoff process and finally (c) a quasi-one dimensional dynamic analysis to model the rapid contact line motion and pinchoff process. In this last stage, we employ a moving contact line model with a dynamic contact angle.

The paper is organized as follows: the experimental setup is depicted in Sec.~\ref{sec:experimental_setup}. The equilibrium (quasi-static) and accompanying stability analysis is stated in Sec.~\ref{sec:static_stability}, while the dynamic model and relevant boundary conditions are described in Sec.~\ref{sec:numerical_model}. Experimental results are presented in Sec.~\ref{sec:rod_experiment}. The stable state of the liquid bridge with a moving contact line is determined in Sec.~\ref{sec:stability_analysis}. Numerical calculations are shown and discussed in Sec.~\ref{sec:numerical_simulation}.

\section{Experimental Setup}\label{sec:experimental_setup}
The experiment setup (Fig.~\ref{fig:geometry}) is similar to the one in our previous study ~\cite[]{qian2009}, but modified for dispensing from a liquid bridge of constant volume. The liquid used is 85/15 (volume) glycerin/water (Sigma-Aldrich) mixture which has viscosity $\mu=84$\,cp, measured with a rheometer (model AR2000N, TA Instruments), at 23\,$^\circ$C, density $\rho=1.222\times10^6\,\rm{g/m^3}$ and surface tension $\gamma=63\,\rm{g/s^2}$ reported from literature~\cite[]{crc2009}. The substrate is a piece of smooth glass slide (VWR plain micro slide) cleaned with piranha solution and coated with a monolayer of octadecyltrichlorosilane (Sigma-Aldrich), on which the liquid exhibits a static receding contact angle $\theta_r=85^\circ\pm 1^\circ$ with an angle hysteresis of 10$^\circ$. Limited experiments were conducted on a less hydrophobic substrate. Care was taken to avoid contamination of the system so as to minimize possible surfactant absorption which might lead to Marangoni stresses.

A small volume of liquid (typical volume: 0.2\,$\mu$L) is deposited on the substrate using a hollow syringe which is connected by a tube to a 10 cc barrel to maintain a constant hydrostatic head at the syringe tip. Bringing the wetted syringe in contact with the substrate and then raising it leaves a small liquid drop on the substrate. The volume of the loaded liquid drop can be controlled by adjusting the syringe size and the syringe speed~\cite[]{qian2009}. Once deposited on the substrate, the liquid drop is translated horizontally using a motorized stage so that it is positioned below a solid cylindrical ``dispensing rod'' (stainless steel, diameter 510\,$\mu$m) which is mounted to a 3D motorized stage (model M-111, Physik Instrumente) and is capable of moving at speeds $U=10 - 1000\mu\rm{m/s}$ with sub-micron per second accuracy.  As the rod approaches and touches the liquid drop, a small constant volume liquid bridge is formed. For measurements at one volume value, multiple dispensing are taken on the same spot of the substrate so that the liquid only needs to be loaded once, which reduces the effect of potential volume variations induced by liquid loading. The size of the liquid bridge is monitored during the data acquisition and it was confirmed that the maximum volume variation due to evaporation was less than $1\%$. 

All experiments were conducted within an air-conditioned room at 23$\pm$0.1\,$^\circ$C. The dispensing setup was positioned on an optical table with tuned damping (model RS4000, Newport) to isolate external vibrations. A high-speed camera (Photron Fastcam APX) equipped with a 5X Mitutuyo lens was used to capture the evolution of liquid bridges at frames rates up to 10 kfps, with a resolution of 3.33\,$\mu$m/pixel. The recorded images were analyzed using MATLAB.

In our experimental system, there are three governing dimensionless numbers: Weber number, Bond number and capillary number. These three dimensionless groups represent the relative importance of inertial force/surface tension, gravity/surface tension and viscous force/surface tension respectively. The small Bond number, ${\rm{Bo}}\equiv{g\rho R^2}/{\gamma}\sim {\cal O}(10^{-2})$, indicates that gravity is insignificant for our experiments. Since the Weber number and the capillary number are defined based on a characteristic speed and the liquid-bridge stretching consists of two phases (Fig. 2a), a proper characteristic speed must be chosen to predict the dominant forces in each phase of bridge stretching. During the initial stretching, the flow speed inside the bridge is of the order of the rod speed $U$. Based on the rod speed, the Weber number and the capillary number are defined as ${\rm{We}_r}\equiv{\rho U^2 R}/{\gamma}\sim {\cal O}(10^{-6})$ and ${\rm{Ca}_r}\equiv \mu U/\gamma\sim {\cal O}(10^{-3})$ respectively. The small $\rm{We}_r$ and $\rm{Ca}_r$ indicates that in the initial stretching only surface tension plays a role. Therefore, the bridge shape is considered to be in equilibrium at each instant of time and the stretching can be treated quasi-statically. However, in the later phase, the liquid bridge retracts rapidly and the capillary wave speed, $u_{cp}=\sqrt{\gamma /\rho R}$, is a more appropriate measure of the flow speed. This choice of the characteristic speed leads to ${\rm{We}_{cp}}\equiv{\rho u_{cp}^2 R}/{\gamma}\sim {\cal O}(1)$ and ${\rm{Ca}_{cp}}\equiv \mu u_{cp}/\gamma\sim {\cal O}(1)$, which reveals that both inertial force and viscous force are comparable to the surface tension and that a dynamic model must be used to appropriately capture the bridge breakup. Note that the capillary number defined with the capillary wave speed is equivalent to the Ohnesorge number.


\section{Theoretical Considerations}

\subsection{Stability of an equilibrium static liquid bridge} \label{sec:static_stability}
Based on the coordinate system defined in Fig.~\ref{fig:geometry}, the static equilibrium profile of an axisymmetric liquid bridge is described by the Young-Laplace equations,
\begin{subeqnarray}
       &r''(s)=-z'(s)\beta '(s), \\
       &z''(s)=r'(s)\beta '(s), \\
       &\beta '(s) = -z(s) + \frac{P-P_0}{\sqrt{\rho g \gamma}} -\frac{z'(s)}{r(s)} \label{eq_shapeLB}
\end{subeqnarray}
with appropriate initial conditions $r(0), r'(0) = \rm{cos}(\theta) ,z(0)=0, z'(0) = \rm{sin}(\theta), \beta(0) = \theta$. Here $s$ is the arc length of the free surface with its origin on the substrate and a prime denotes differentiation with respect to the arc length. $r(s)$ is the radius of the liquid bridge, $z(s)$ is the vertical distance from the substrate, and $\beta(s)$ is the angle between the radial axis and the tangent to the interface. $\rho$ is the liquid density, $\gamma$ is the surface tension, and $g$ is acceleration of gravity. The pressure difference $(P-P_0)$ across the liquid interface at the coordinate origin is adjustable to make the solution satisfy the boundary condition of $r(s^*)=R$, in which $s^*$ is the point at the rod $(z=h)$. For a liquid bridge having a volume $v$ and a height $z(s^*)=h$, only one of the two initial conditions: the contact line radius $r(0)$ and the contact angle $\theta$ should be specified to determine the shape and the other is a free parameter to fulfill the volume constraint: $\int^{h}_0 \pi r^2 dz=v$. Which initial condition is to be specified depends on the motion state of the contact line: pinned or receding.

The stability of an equilibrium liquid bridge can be determined according to the method introduced by \cite{Myshkis}, which is an eigenvalue problem
\begin{subeqnarray}
       &L\varphi_0 + \nu = \alpha\varphi_0   \quad(0\leq s \leq s^*) \\
       &\varphi_0(0)=0, \varphi_0(s^*) = 0, \int_0^{s^*}r\varphi_0ds=0,\\
       &L\varphi_1 + \frac{1}{r^2} = \alpha\varphi_1  \quad(0\leq s \leq s^*),\\
       &\varphi_1(0)=0, \varphi_1(s^*) = 0,
\end{subeqnarray}
Here
\begin{subeqnarray}
       &L\varphi\equiv -\varphi ''-\frac{r'}{r}\varphi'-a(s)\varphi,\\
       &a(s) = -r'(s)-{\beta '}^2(s) - \left(\frac{z'(s)}{r(s)}\right)^2 \label{eq_stability}
\end{subeqnarray}
$\nu$ is an unknown constant beforehand, and primes denote derivatives with respect to $s$. $\varphi_0$ and $\varphi_1$ correspond to the axisymmetric perturbations and the most dangerous nonaxisymmetric perturbation to the liquid bridge respectively. $\alpha$ is the eigenvalues of Eq.~\ref{eq_stability}. A positive sign of the smallest eigenvalue signifies a stable equilibrium liquid bridge and a vanishing smallest eigenvalue represents the critical state of a liquid bridge. The stability boundary for a bridge with given contact line radius, $r(0)$, can be computed following the numerical algorithm described in \cite[]{slobozanin1993} and then the stability of the calculated liquid bridge is known.

The liquid bridge is assumed to stretch through a sequence of quasi-static states. At each state, the bridge height is specified and the stable equilibrium bridge profile can be solved. Successively applying the calculation for varying heights by a small step, we can track the evolution of the bridge. This tracking process is terminated when the bridge stretches to a critical height at which no equilibrium solution exists or the solution becomes unstable. The critical height can be pinpointed by iterating with a refined height step.

\subsection{Numerical model of a dynamic stretching liquid bridge} \label{sec:numerical_model}
Beyond the critical equilibrium state, the bridge starts breaking and its shape deforms quickly. This is no longer quasi-static motion. To simulate the dynamic breakup, the axisymmetric Navier-Stokes equations need to be solved with appropriate kinematic and traction boundary conditions~\cite[]{yildirim2001}. To simplify the mathematical treatment, ~\cite{eggers1994} used the slender-jet approximation to truncate the high order terms in the 2D governing equations to arrive to a set of 1D model equations,
\begin{subeqnarray}
       &\partial_t{r} + ur_z=-ru_z/2,\\ \label{eq_mass}
       &(\partial_t{u} + uu_z)=-\frac{\gamma}{\rho}\kappa_z +\frac{3 \mu}{r^2}[(r^2u_z)_z]-g , \label{eq_momentum}
\end{subeqnarray}
Here $u(z,t)$ and $r(z,t)$ are the axial flow speed and column radius. $g$ is acceleration of gravity. The full axisymmetric mean curvature, $\kappa$,
\begin{equation}
       \kappa=\frac{1}{r(1+r_z^2)^{1/2}} - \frac{r_{zz}}{(1+r_z^2)^{3/2}} . \label{eq_curvature}
\end{equation}
is maintained to precisely predict the bridge shapes. This model has been successfully applied to studying jet breaking ~\cite[]{eggers1994} and liquid bridge stretching~\cite[]{zhang1996,qian2009}. Comparisons in the numerical results for stretching bridges with fixed contact lines, $r(0)/R=1$, between the exact 2D model and the approximated 1D model showed that the 1D model gives an accurate prediction of the macroscopic features of the bridge breakup as long as the ratio of the stretching speed, $U$, to the capillary wave speed, $u_{cp} = \sqrt{\gamma / \rho R}$, is much less than one~\cite[]{yildirim2001}. In our experiment, a typical capillary wave speed is 10\,cm/s, which is much larger than the used stretching speeds. Although close to the bridge pinch-off the high radial speed violates the assumption of the 1D model, we nevertheless use this model to simulate the dispensing process, and will discuss its accuracy and limitations.  As mentioned earlier, the Bond number in our experiments is very small.  Nevertheless, we retain the gravity term in the system of equations for generality.  We also note that this model reduces to the Young-Laplace equation (eq. \ref{eq_shapeLB}) in the limit of quasi-steady motion and can be used without restriction to model the entire bridge evolution history.

To solve the 1D equations (\ref{eq_mass} - \ref{eq_curvature}), boundary conditions at both ends of the bridge should be specified. At the top of the liquid bridge, the contact line is pinned, $r(h)=R$, and no flow penetrates the rod surface, $u(h)=0$. On the substrate, the axial flow speed is zero, $u(0)=0$. Since the contact line allows to move freely, $r(0,t)$ is unknown and the contact angle $r_z(0,t)=\rm{cot}(\theta)$ must be prescribed. The simplest way to model the contact angle is to define $\theta(t)$ as a constant. However, the fixed contact angle model is not able to capture the contact angle dependence on the contact line speed, $u_c=r_t(0,t)$, ~\cite[]{Oron1997,bonn2008}. An improvement, thus, is to relate the dynamic contact angle to the contact line speed using an empirical equation ~\cite[]{tanner1979,dussan1979}
\begin{equation}
   u_c = \lambda (\theta-\theta_r)^n. \label{eq_tanner}
\end{equation}
Here ($\theta-\theta_r$) is the deviation of the dynamic contact angle from the static receding angle. $\lambda$ is an empirically determined constant that is a measure of the contact line speed. $n$ is another empirical constant which was experimentally observed to be between 1 and 3~\cite[]{dussan1979}. This dynamic contact angle model has been successfully applied to simulate the spin coating ~\cite[]{wilson2000} and pin-tool printing~\cite[]{Panditaratne_PHD2003}. In this paper, we adopted this model as a boundary condition into the numerical calculation.

\section{Results and Discussion}

\subsection{Overall behavior of drop deposition from constant-volume liquid bridges} \label{sec:rod_experiment}

\begin{figure}
\begin{center}$
\begin{array}{cc}
\includegraphics[width=4.5in]{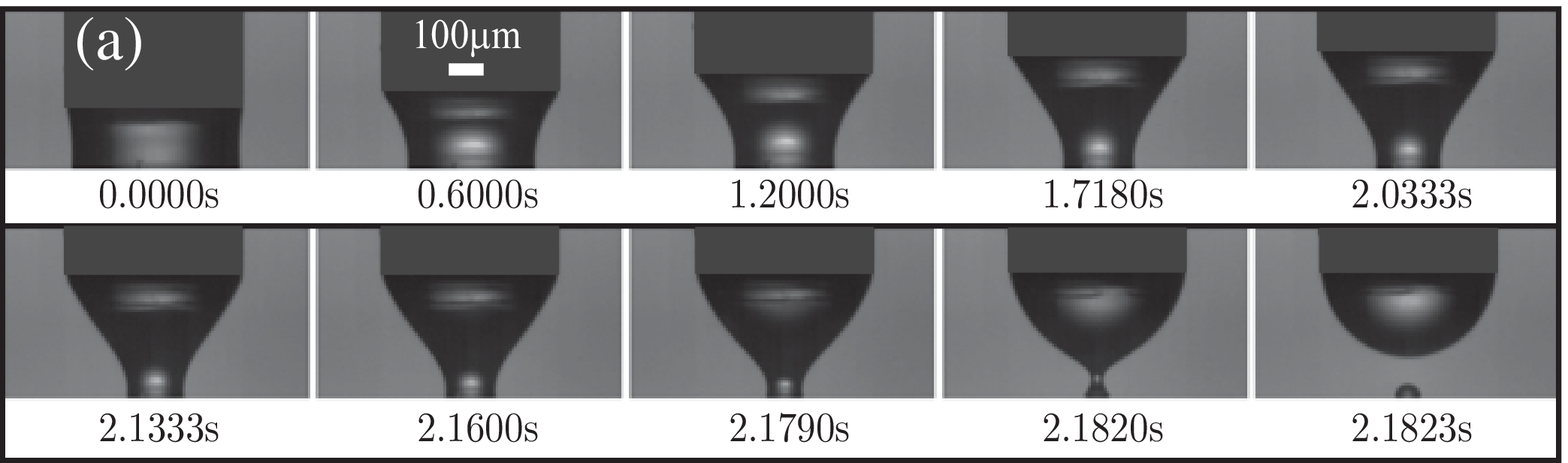} \\
\includegraphics[width=3.2in]{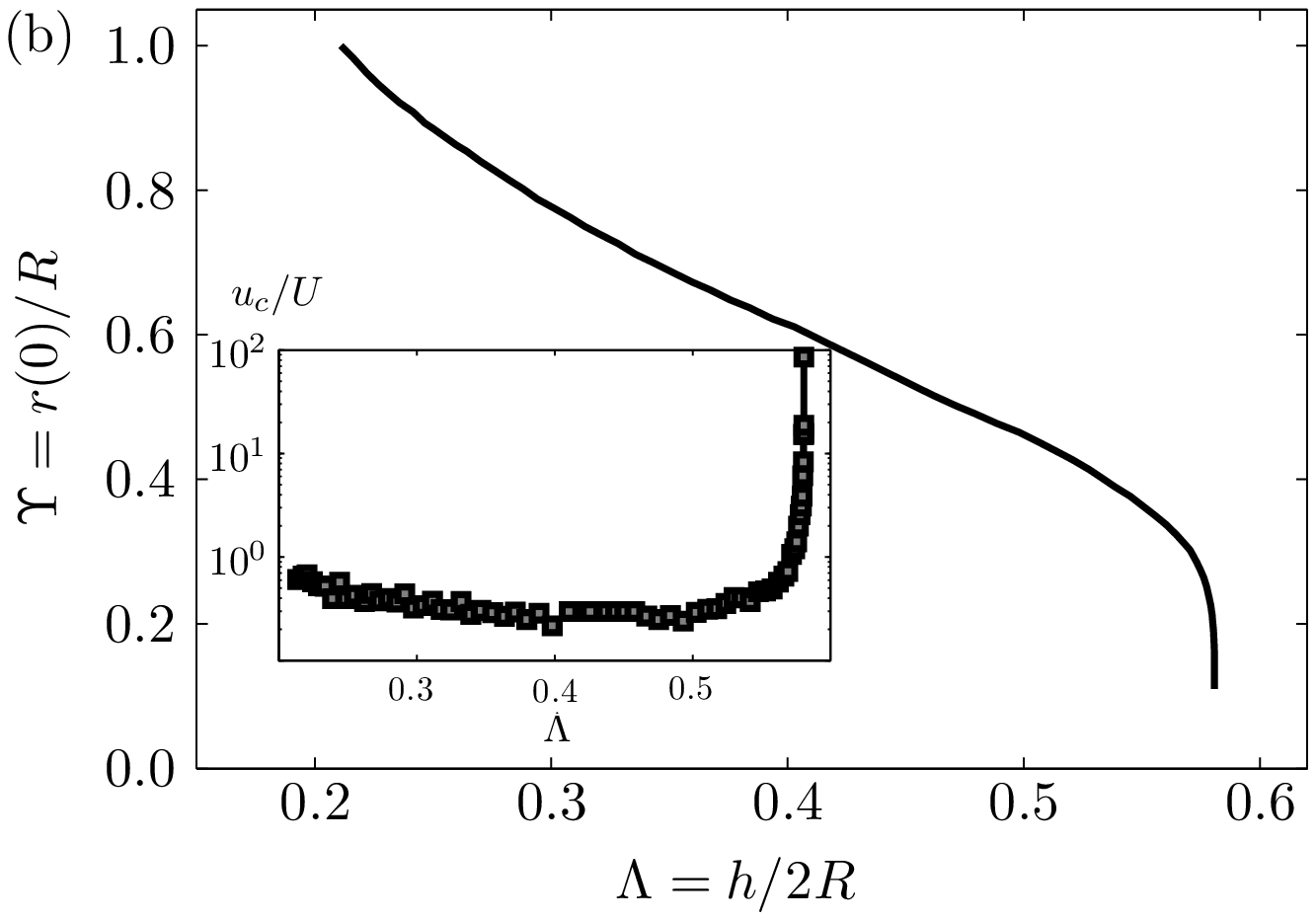}
\end{array}$
\end{center}
\caption{(a) Typical sequence of images of drop dispensing on a hydrophobic surface with a receding contact line. Two stages of contact line motions were observed: a slow retreating at the beginning (top row) and a rapid retraction prior to the bridge pinch-off (bottom row). The rod is 510\,$\mu \rm{m}$ in diameter and it lifts at a constant speed 100\,$\mu \rm{m/s}$. The gray intensity of the rod is changed for contrast to the liquid bridge. (b) Corresponding measured contact line locations $\Upsilon=r(0)/R$ as a function of bridge's heights $\Lambda=h/2R$. The inset shows the contact line speeds $u_c/U$ measured from the evolution of the contact line.}
\label{fig:exp_img}
\end{figure}

\begin{figure}
\centering
\includegraphics[width=4.5in]{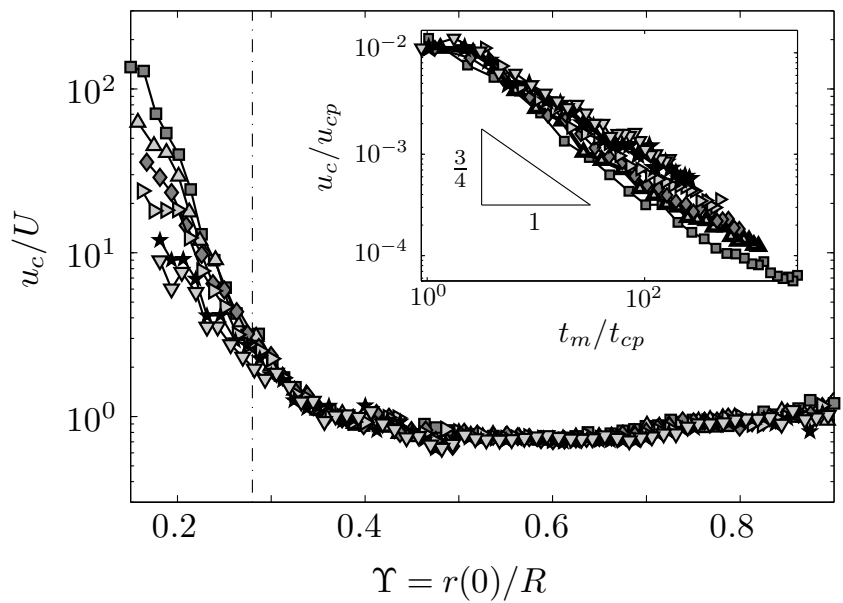}
\caption{Experimentally measured non-dimensional contact line speeds from a stretching constant-volume bridge for $U = 35 (\Box), 80 (\bigtriangleup), 125 (\diamond), 200 (\triangleright), 400 (\bigstar)$ and $ 600 (\bigtriangledown) \,\mu \rm{m /s}$. The dash-dot line denotes the predicted critical radius from static analysis, at which the liquid bridge becomes unstable and starts breaking. The inset shows the dimensionless contact line speed as a function of time to the breakup, $t_m$. Speed and time is scaled by $u_{cp}=\sqrt{\gamma/\rho R}$ and $t_{cp}=\sqrt{\rho R^3/\gamma}$.} %
\label{fig:cl_speed_exp}
\end{figure}

The general behavior of the drop deposition experiment is reviewed here, and summarized graphically in Fig.~\ref{fig:exp_img}. As the rod retracts (and the bridge height, $\Lambda= h/2R$, increases), the liquid bridge stretches, developing axial curvature (and hence negative pressure inside the liquid volume). The measured receding contact line speed, $u_c/U$, versus the contact line radius, $\Upsilon = r(0)/R$, is shown for several retraction speeds, $U$, in Fig.~\ref{fig:cl_speed_exp} (for convenience, the contact line speed is defined as positive when the contact line contracts). As discussed earlier, we identify two phases of the bridge evolution. Initially, the contact line moves at a low speed which is comparable to the stretching speed, $u_c/U\approx 1$. Below a critical radius the contact line accelerates quickly to a speed comparable to the capillary wave speed . Although this behavior is generically similar to that observed in constant-pressure deposition~\cite[]{qian2009}, we see two chief differences. Firstly, during the low-speed phase of constant-pressure deposition, the normalized contact line speed, $u_c/U$, shows a weak dependence on $U$, whereas in the constant-volume case the speed collapses with no further dependence on the stretching speed. This scaling confirms the assumption of quasi-static stretching during this phase, since the contact line location, $r(0)$, as well as the bridge shape, is solely determined by the bridge height, $h$. In the constant pressure case, the weak dependence on $U$ is due to the fact that the bridge volume increases with time. Secondly, in the constant-pressure case, both the critical radius at which the contact line starts to accelerate and the radius at which the contact line reaches its maximum speed decrease with the retraction speed, $U$. In contrast, for the current case, there is no discernible change in the critical radius as a function of $U$, and the variation in the maximum contact line speed location is more moderate than was observed in the constant-pressure case. 

In the rapid-retraction phase, plotting the dimensionless contact line speed (inset of Fig.~\ref{fig:cl_speed_exp}) as a function of the time to bridge breakup, $t_m$, shows a power-law dependence, $u_c/u_{cp}\sim (t_m/t_{cp})^{-3/4}$. Although we have no physical explanation for it, this power-law dependence is also predicted by numerical calculations and its exponent is found to be dependent on the parameters of Tanner's law (Eq.~\ref{eq_tanner}). Similar power-law behaviors have been observed (also without satisfactory explanation) in drop coalescence and wetting~\cite[]{bonn2005,siggi2005,bird2008} who also found that the scaling exponent depends on the type of force resisting drop deformation~\cite[]{stone1999} and the static equilibrium contact angle~\cite[]{bird2008}. Note that the speed and time scale, $u_{cp}=\sqrt{\gamma/\rho R}$ and $t_{cp}=\sqrt{\rho R^3/\gamma}$, are determined only by the liquid properties which are fixed in the experiment and thus the dimensional maximum contact line speed weakly relies on the dispensing speed. However, the dimensional contact line speed at the starting acceleration point does change with $U$ and it is approximately equal to $U$. Therefore, the time taken to accelerate to the maximum contact line speed, and the distance that contact line retreats within that time both decrease with $U$. Combined with the fact that the critical radius is weakly dependent on the dispensing speed, we can conclude that the deposition drop size should increase with the dispensing speed for deposition from a constant-volume bridge.

\subsubsection{Drop size}
\begin{figure}
\centering
\includegraphics[width=4.5in]{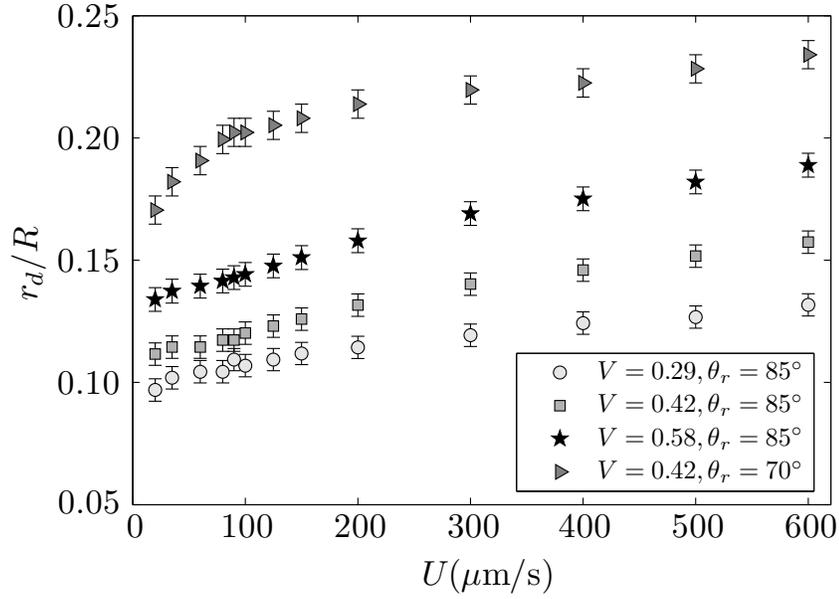}
\caption{Experimentally measured dispensing drop sizes vary with rod speeds for three bridge volumes, $V=v/\pi R^3$. The liquid bridges have a static receding contact angle $\theta_r=85^\circ$ on the surface.  Also shown are drops dispensing on a less hydrophobic surface $\theta_r=70^\circ (\triangleright)$ for one volume value $V=0.42$. The rod has a diameter ($2R$) of 510\,$\mu$m.}
\label{fig:rod_drop}
\end{figure}

This expected dependance of the drop size on the retraction speed is confirmed in Figure~\ref{fig:rod_drop}, which shows the change in the dispensed drop size as a function of the rod speed for several different liquid volumes. As argued above, for a given volume, the drop size increases as the rod speed increases. This contrasts to that observed in constant-pressure deposition, in which drop sizes dramatically decreases with increasing syringe speed, reaching a minimum drop size, after which $r_d$ starts to increase slowly ~\cite[]{qian2009}. Comparison of drop sizes between different bridge volumes reveals that increasing the volume causes the drop size to increase, and furthermore, that the increase with $U$ is more apparent for the larger initial volumes. Additionally, some data (some of which is shown in Fig.~\ref{fig:rod_drop}) from experiments using a less hydrophobic surface ($\theta_r=70^\circ\pm1^\circ$ with a contact angle hysteresis of 15$^\circ$) shows that the drops generated on the less hydrophobic surface are larger than on a more hydrophobic surface for the same bridge volume. Unfortunately, difficulties in preparing consistent surfaces with a variety of contact angles prevented us from systematically investigating the effects of surface wettability on the resultant drop size.

\subsection{Stability of static liquid bridge} \label{sec:stability_analysis}
Having described the overall behavior of the drop deposition, the following sections use the analytical methods described earlier to quantitatively model the details of both the static, and dynamic phases of the process. In this section, we use the calculation procedure, described in sec.~\ref{sec:static_stability}, to predict the height at which the contact line first moves inward, and the critical contact line radius at which the bridge becomes unstable.

\subsubsection{The first movement of a contact line}
\begin{figure}
\begin{center}
\includegraphics[width=4.5in]{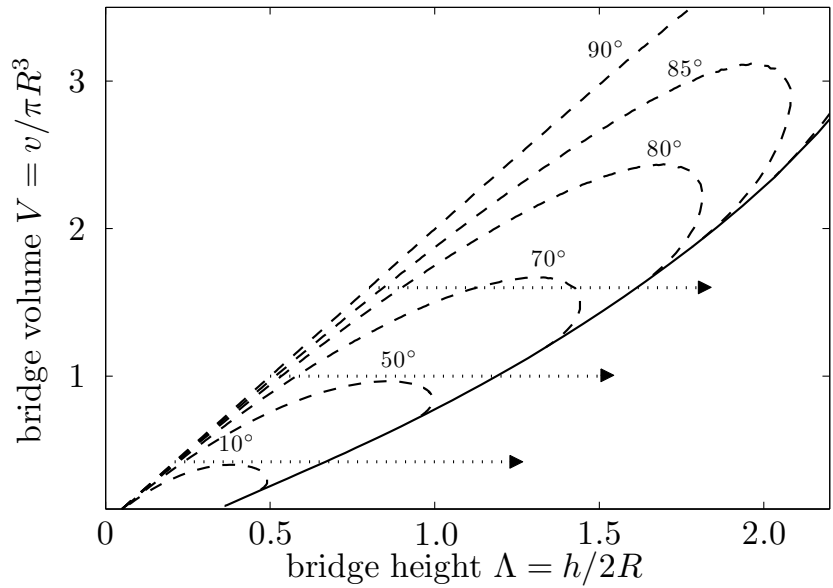}
\caption{Contact angles as a function of bridge height, $\Lambda$, and volume, $V$. A typical experiment for a given fixed bridge volume is represented by a horizontal dotted arrow and illustrates the  change in contact angles experienced as the bridge is extruded.  This example is for a fixed contact line position, $\Upsilon = r/R = 1$.  The solid curve to the right of the frame represnts the static stability boundary for a liquid bridge. }
\label{fig:stability_fix}
\end{center}
\end{figure}

The stable state of a liquid bridge with a fixed contact line can be determined in a plane of dimensionless bridge height, $\Lambda=h/2R$, and volume,  $V=v/\pi R^3$ (Fig.~\ref{fig:stability_fix}). Initially, the liquid bridge is cylindrical, with the contact line at $\Upsilon = r(0)/R = 1$ and the contact angle at 90$^\circ$, represented by the left-most contour. For a fixed bridge volume, the experimental evolution follows a horizontal line,  starting at this left-most line, and moving towards the right (indicated by the dotted arrows). As the rod retracts, the liquid bridge stretches and the contact angle at the substrate changes, first decreasing, reaching a minimum, and then increasing again.  However, the bridge cannot be extruded indefinitely, and the stability boundary at which a statically stable bridge can no longer exist (Eq.~\ref{eq_stability}) is shown as the solid boundary on the right side of Fig.~\ref{fig:stability_fix}.  Crossing this boundary marks the onset of the rapid retraction and pinchoff phase.

However, the example shown in figure~\ref{fig:stability_fix} is only valid for a given contact line position, $\Upsilon$, in this case, a bridge whose contact line is pinned at $\Upsilon=1$.  On a flat, smooth substrate as $\theta$ decreases, it may cross the contour corresponding to the static receding contact angle, $\theta_r$ and the contact line will start to retreat.  At this point, a new series of $\Lambda-V$ maps, each representing the instantaneous value of $\Upsilon$ need to be consulted.

\subsubsection{Evolution of freely moving contact lines}
\begin{figure}
\begin{center}
\includegraphics[width=4.5in]{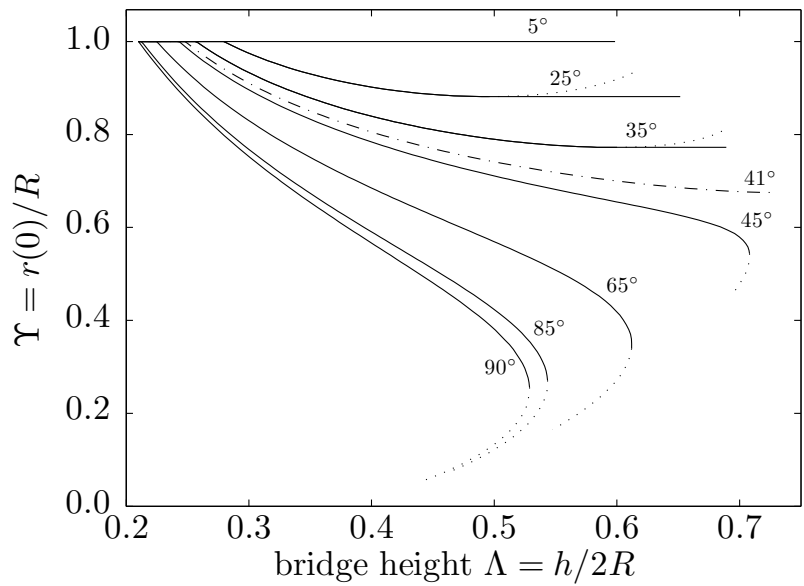}
\caption{ Static calculation of the locations of a free moving contact line $\Upsilon$ as a function of the bridge height $\Lambda$ under a contact angle constraint $\theta\geq\theta_r$. Calculated evolution curves (solid line) are shown for different limiting wetting angles $\theta_r$. The dotted line shows the theoretically possible and stable but not practically feasible contact line locations. The dash-dot line shows the boundary across which the transition takes place from the fixed-receding-fixed contact line to the fixed-receding contact line. } 
\label{fig:evl_path}
\end{center}
\end{figure}

As the contact line moves, it is more convenient to analyze the problem in terms of the contact line position, $\Upsilon = r(0)/R$, versus the bridge height, $\Lambda$. For a given volume in the liquid bridge, the limiting wetting angles define three types of contact line behavior, and they can be identified on the ($\Lambda,\Upsilon$) plane: always-fixed contact line, fixed then receding contact line, and fixed-receding-fixed contact line (Fig.~\ref{fig:evl_path}). For a low limiting wetting angle, for example $5^\circ$, the contact angle never falls below the critical value, and the contact line is always pinned at its initial position $\Upsilon = 1$ until the bridge becomes unstable. During the stretching, the contact angle decreases initially and then increases again (as predicted by Fig~\ref{fig:stability_fix}).  

For large minimum wetting angle, such as $\theta = 65^\circ$, the contact line is initially fixed, but as $\Lambda$ increases, the contact angle reaches its limiting wetting angle and the contact line start to recede. The radius of the
contact line continues to decrease monotonically with $\Lambda$ (with the contact angle at its limiting value).  However, at a critical height, there exists no statically stable solution (the evolution path becomes vertical), and the bridge breaks. Note that for a given contact angle, $\theta$, there exists a second static equilibrium solution (dotted lines) for the same height $\Lambda$, but with a  different contact line radius, $\Upsilon$. However, in our experiments the contact line recedes smoothly and no``radius jumping'' was observed, so the lower branch of the  static solution (dotted line) was not practically accessible.

For intermediate limiting wetting angles, for example, $\theta = 35^\circ$, the initial contact line behavior is as before: fixed initially, then retreating. However, for these cases, the theory predicts that when the bridge elongates to a certain height, the contact line reaches a minimum radius and then starts to expand (dotted line). In practice,
however, contact angle hysteresis will result in a different observed path, and the contact line will stop moving while the contact angle grows above the receding angle (a horizontal line on the $\Lambda - \Upsilon$ plane). One might then observe a situation in which the contact angle grows to its limiting advancing value, at which point the contact line would then expand, now on a different $\theta$ line.

In our experiment, the liquid on the substrate exhibited a high static receding contact angle ($85^\circ\pm 1^\circ$) and so only the fixed-receding type of contact line behavior was observed. From the Figure~\ref{fig:evl_path}, it is clear that at a high limiting wetting angle the evolution path has a steeper slope and the contact line radius, $\Upsilon$, reduces faster with the stretching height. The end point of a evolution path denotes the critical radius at which the bridge breaks. For the fixed-receding contact lines, the breaking height and the critical radius both increase with the reducing wetting angle.

\begin{figure}
\begin{center}
\includegraphics[width=5.4in]{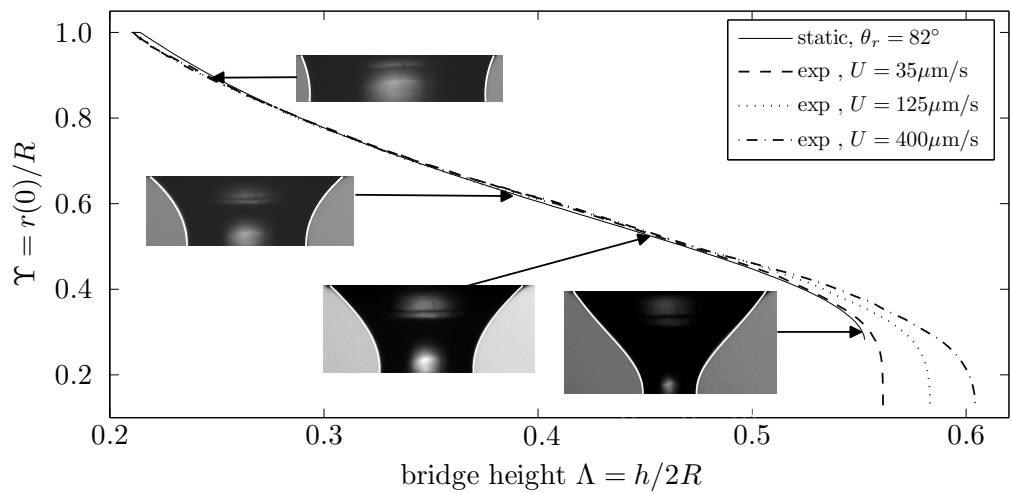}
\caption{Evolution of contact lines from static calculation (solid) compared to those from experiments at three different stretching speeds $U$ (dashed) . The insets show the comparison of calculated bridge shapes (white line) to that imaged from experiment at a stretching speed $U=35\mu\rm{m/s}$.  } 
\label{fig:evl_path_exp}
\end{center}
\end{figure}

The prediction for $\theta_r=82^\circ$ is compared with our experimental data (Figure~\ref{fig:evl_path_exp}) for three retraction speeds.  The angle chosen for calculation is slightly lower than the static receding angle ($85^\circ$) observed in experiment due to the fact that the dynamic contact angle is always a little lower than the static angle. The dynamic contact angle is velocity-dependent \citep[]{dussan1979,tanner1979,bonn2008} and it varies during the bridge stretching. However, during the quasi-static phase, the contact line speed is so small that the variations in the dynamic contact angle variation are too small to significantly affect the evolution path, and the static theory gives an excellent prediction of the contact line motion and the overall bridge profile for all three rod speeds. Close to the critical state, the contact line speed is much faster and the dynamic contact angle decreases more noticeably from its equilibrium value. This behavior results in an increase of the breaking height with higher rod speed. Additionally, as the rod speed increases, the viscous forces inside the liquid bridge are larger, which help to stabilize the stretching bridge and further postpone the breakup ~\cite[]{zhang1996}. For both these reasons, the agreement between the experiment and the simple static theory is compromised in this region.

\subsubsection{Effects of volumes and minimum wetting angles on the critical radius}
\begin{figure}
\centering
\includegraphics[width=4.5in]{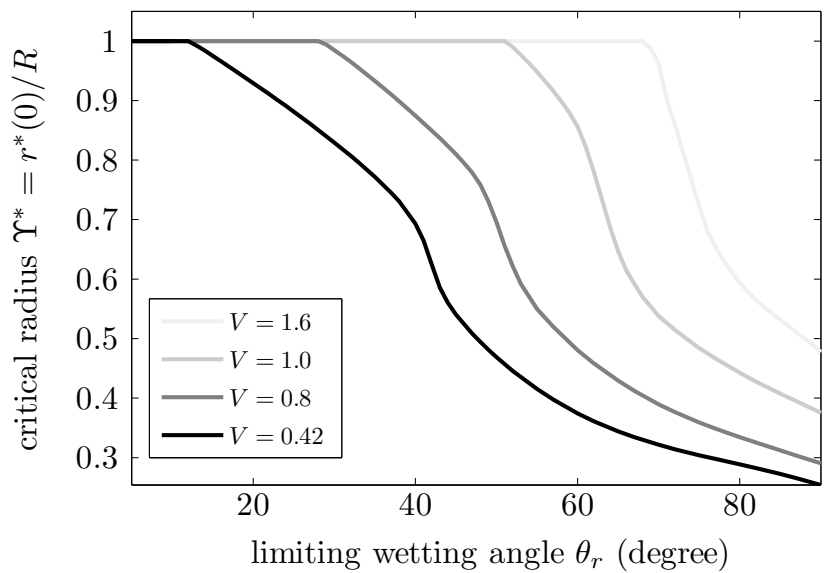}
\caption{Calculated critical radius $\Upsilon^*=r^*(0)/R$ as a function of limiting wetting angles $\theta_r$ from static theory. Calculation are shown for four liquid volumes $V=0.42, 0.6, 1.0, 1.6$. }
\label{fig:unequal_contactangle}
\end{figure}

The critical radii at which the contact line starts to move, defined as $\Upsilon^*$, are shown as a function of the minimum wetting angles for four different bridge volumes in Figure~\ref{fig:unequal_contactangle}. For each volume, there exists a wetting angle below which the critical radius remains at $\Upsilon^*=1$, meaning that contact angle never reaches the minimum wetting angle and the contact line does not retreat during the bridge stretching process. For these fixed contact line cases, the neutrally-stable state of the liquid bridge is determined only by the bridge volume and height ($V,\Lambda$) \cite[]{slobozanin1993}. However, if the wetting angle is higher than this threshold, the critical radius monotonically decreases as the limiting wetting angle rises. Since the higher limiting wetting angle is equivalent to lowering the surface wettability, the bridge starts breaking at a smaller value of $\Upsilon$ on a more hydrophobic surface, which partially accounts for smaller deposition drops (Fig.~\ref{fig:rod_drop}). Moreover, the comparison in critical radius between different values of the bridge volume shows that, for the same minimum wetting angle, the smaller the volume, the smaller the critical radius that can be reached. For constant-pressure deposition, increasing the stretching speed is equivalent to lowering the rate at which the volume increases, and thus decreasing the critical radius. Therefore, raising the syringe speed creates smaller drops until the syringe speed is much higher compared to the speed of fluid into the drop, at which the volume variation is negligible and the effect of the dynamic contact angle variation becomes dominant.

\subsection{Dynamics of the stretching liquid bridge} \label{sec:numerical_simulation}
The static theory discussed above yields good predictions of the evolution of the liquid bridge prior to the critical state but, not surprisingly, it fails beyond this equilibrium boundary. To accurately model the dynamic bridge breakup and the contact line motion after the critical state, numerical calculations were performed using the one-dimensional model described in sec.~\ref{sec:numerical_model}. Both boundary conditions - the fixed contact angle and the dynamic variable contact angle - were implemented in numerical simulations to explore the effects of the velocity-dependent dynamic contact angle on the drop dispensing process.

\subsubsection{Predicting drop size}\label{sec:drop_size_numl}
Figure~\ref{fig:rod_drop_tanner} compares the calculated final drop sizes (solid line) for different rod speeds with those obtained from experiments (symbols) for three bridge volumes. The dynamic variable contact angle model was adopted in the numerical calculation, and the model parameters were adjusted to obtain good agreement between theory and experiment. Given this, it is perhaps not surprising that the numerical results agree well with the experiments, although it is reassuring that using a single set of parameters ($\theta_r = 85^\circ,\lambda = 0.02,n=1$) we are able to capture both the trend in which the drop size increases with rod speed, as well as the quantitative values obtained from experiment. Only at large bridge volumes do the predictions deviate in any significant manner from the measurements. However, even in this regime, the assumptions of the one-dimensional theory should remain valid, suggesting perhaps that the reason for the discrepancy between simulation and experiment is the failure of the simple Tanner law model for the dynamic contact angle. The simulations are extremely sensitive to the contact angle. Using a static contact angle, $\theta_r$, (dashed line) in the calculations gives very poor agreement with the corresponding measurements (squares), predicting drop sizes that are much smaller and only weakly dependent on the rod speed. Although experimental measurements are limited to small bridge volumes, numerical calculations have been run for much higher volume values, $V\leq 2.0$, and the same trend are observed in the relation of drop size to the rod speed.

\subsubsection{Contact line speed}
\begin{figure}
\centering
\includegraphics[width=4.5in]{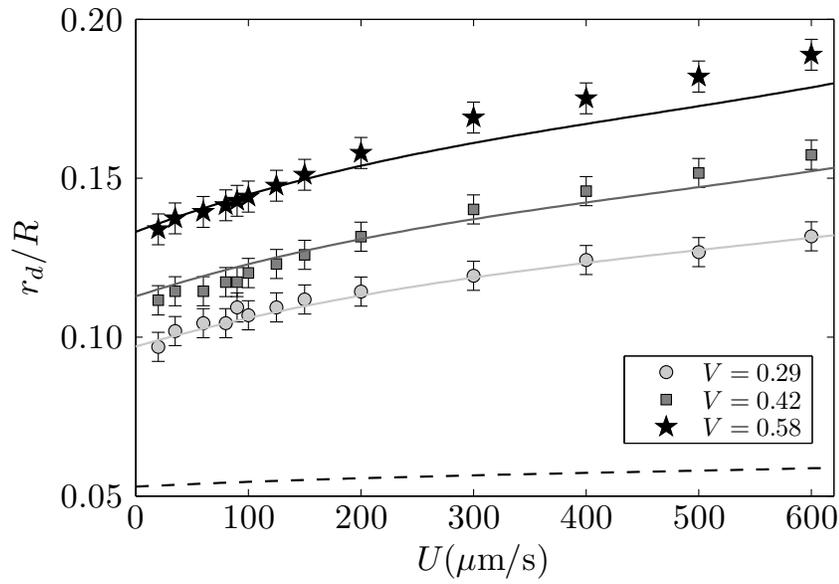}
\caption{Calculated dispensing drop sizes (solid line) compared to experiment (symbols) for three bridge volumes. Here a dynamic contact angle model was included in the calculations with chosen parameters, $\lambda = 0.03 ,n=1,\theta_r=85^\circ$. Also shown are dispensing drop sizes from fixed contact angle calculation with $\theta_r=80^\circ$ for one volume $V=0.42$ (dashed line). }
\label{fig:rod_drop_tanner}
\end{figure}

\begin{figure}
\centering
\includegraphics[width=4.5in]{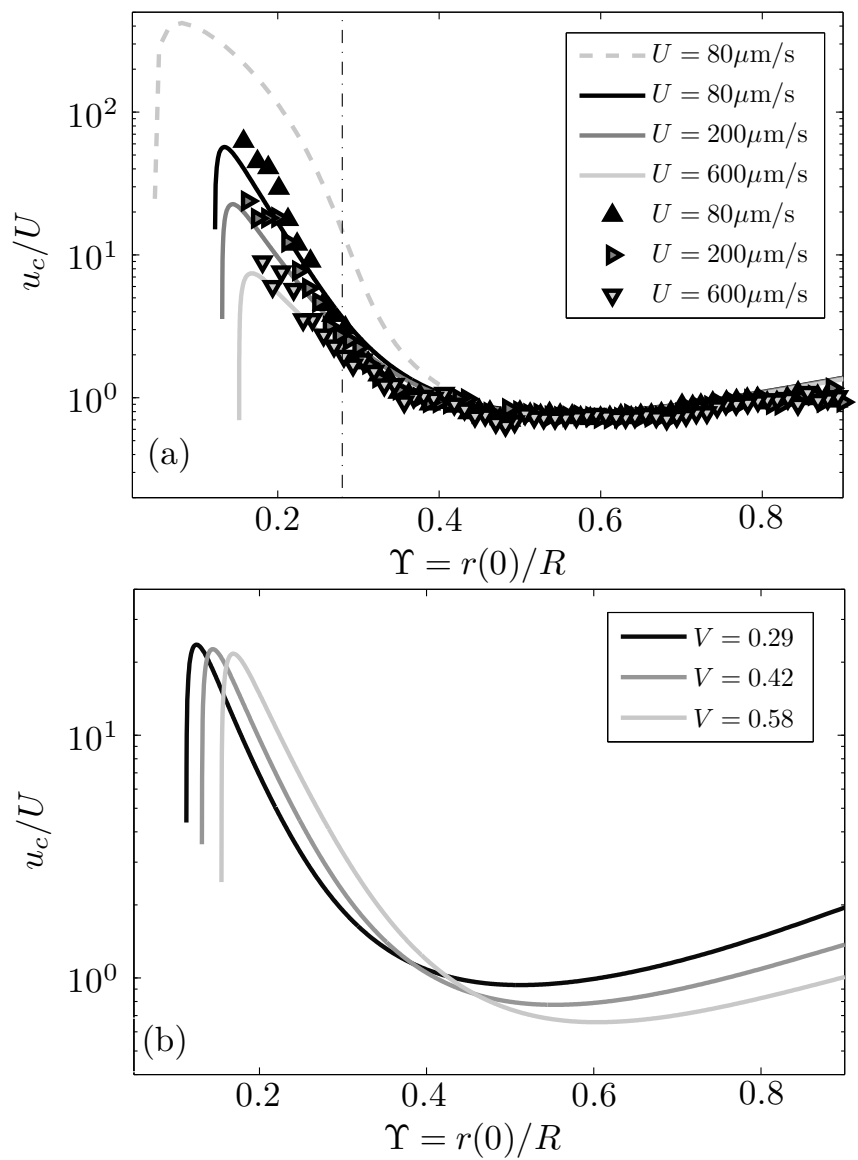}
\caption{(a) Comparison in contact line speeds between experiments (symbols) and numerical calculations with the velocity dependent contact angle model in parameters $\lambda = 0.02 ,n=1,\theta_r=85^\circ$ (solid line) and the fixed contact angle model $\theta_r=80^\circ$ (dashed line) for $V=0.42$. The dash-dot line denotes the predicted critical radius from static analysis. (b) Computed contact line receding speed at one stretching speed $U=200\,\mu \rm{m/s}$ for different volume values. A dynamic contact angle is applied as a boundary condition with parameters $\lambda = 0.02 ,n=1,\theta_r=85^\circ$.  }
\label{fig:rod_cl_numl}
\end{figure}

The discrepancy between the predictions obtained using the fixed and the dynamic variable angle calculations can be explained by the behavior of the contact line motion (Fig.~\ref{fig:rod_cl_numl}-a). The contact line speeds from the dynamic contact angle calculation (solid lines) are plotted along with a single example from the fixed contact angle calculation (dashed line). Both calculations capture the general behavior observed in experiment. In the slowly-receding stage, calculations using both boundary conditions show only slight differences and both show little dependence on retraction speed (due to small variations in the receding angles). However, upon entering the high-speed stage, the contact line with a fixed contact angle moves much faster and the maximum contact line speed is ten times higher than that obtained using the dynamic angle calculation. The higher contact line speed leads directly to a smaller predicted drop size. In contrast, the contact line speed obtained using the dynamic contact angle calculation agrees with the experimental measurement not only in the slowly-receding stage but also in the high speed stage (Fig.~\ref{fig:rod_cl_numl}-a). It is therefore not surprising that the dynamic contact angle model has a good prediction of dispensing drop sizes (Fig.~\ref{fig:rod_drop_tanner}). Note that numerical calculations predict a maximum contact-line speed immediately before bridge breakup. This was not observable in experimental measurements due to the limited imaging rate. Before the maximum contact-line speed is reached, the calculated axial flow speed along the bridge is positive everywhere (pointing upward toward the rod) and therefore the contact line retreats to satisfy the conservation of mass. However, right before bridge pinchoff, a negative flow speed develops between the bridge neck and the solid surface. This reverse flow prevents the contact line from retreating further, leading to a sudden drop in the contact-line speed and subsequently a pinned contact line. Comparing the maximum speeds found at different rod speeds shows that the radius at which contact line speed achieves its maximum increases slightly with $U$, in agreement with the experimental observation and the earlier discussion in sec.~\ref{sec:rod_experiment}.

Also shown (Fig.~\ref{fig:rod_cl_numl}-b) are the contact line speeds for three different bridge volumes, all obtained using the dynamic contact angle calculation. It is obvious that the larger volume liquid bridge experiences the sharp contact line acceleration at an earlier (i.e. larger) value of $\Upsilon$,  and reaches a slightly lower maximum speed.  Both of these contribute to the increase in the resultant drop size as a function of bridge volume (Fig.~\ref{fig:rod_drop_tanner}).

\subsubsection{Effects of surface wettability on drop sizes}
\begin{figure}
\centering
\includegraphics[width=4.5in]{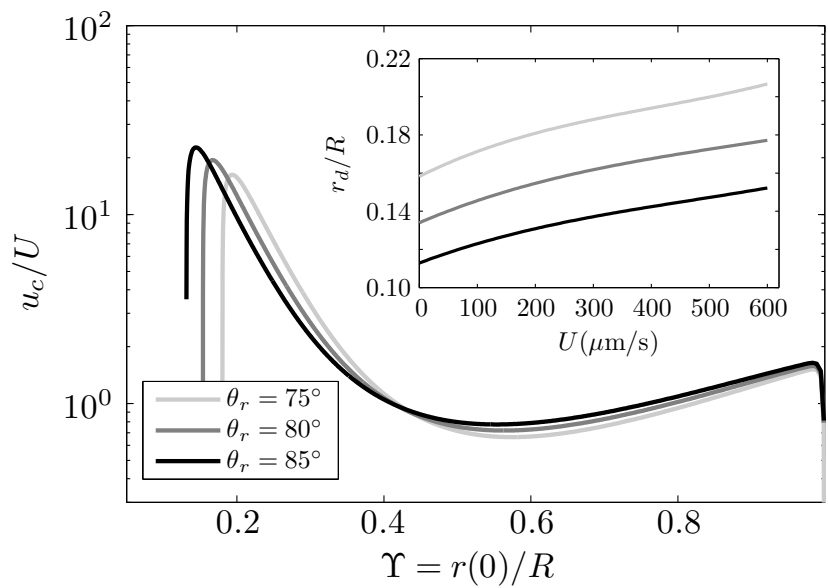}
\caption{Effects of static receding angles $\theta_r$ on the contact line speeds for one stretching speed $U=200\,\mu \rm{m/s}$ from  dynamic contact angle calculation with the model parameters $\lambda = 0.02 ,n=1$. Corresponding dispensing drop sizes vs. stretching speeds are shown in the inset. }
\label{fig:drop_ca_numl}
\end{figure}

The effects of different static receding angles on the dynamics of drop dispensing were not accessible in our experiments due to the difficulty of obtaining high quality surfaces with different surface characteristics. However, this is easily explored in the simulations by changing $\theta_r$ (keeping the other parameters in the model fixed at $\lambda=0.02,n=1$). Figure~\ref{fig:drop_ca_numl} shows the contact line speed, and resultant drop sizes for three static receding contact angles ($V=0.42,U=200\mu\rm{m/s}$). It is clear that as $\theta_r$ decreases (i.e. increasing wettability), the liquid bridge becomes unstable at a larger critical radius, as predicted by the static analysis. In the dynamic pinch-off process, the stronger viscous drag prevents the contact line from following the rapid traction of the liquid bridge, which results in a lower maximum contact line speed and larger resultant drop size (Fig.~\ref{fig:drop_ca_numl}). The same effects on the contact-line motion and the resultant drop size are observed (not shown) for $\theta_r>90$ in calculations.

\subsubsection{Sensitivity of drop sizes to the model parameters}
\begin{figure}
\centering
\includegraphics[width=4.5in]{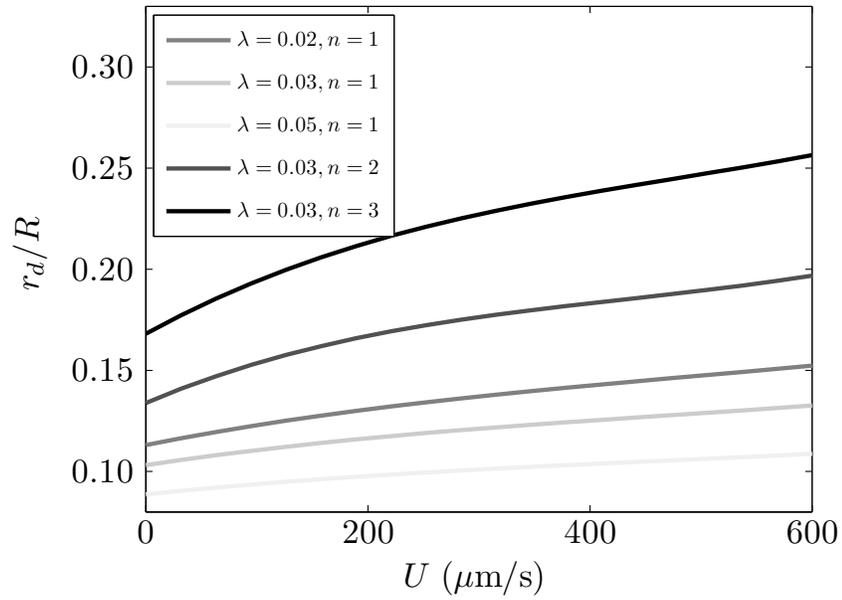}
\caption{Influences of model parameters on the numerically calculated deposited drop sizes from a constant-volume liquid bridge, $V=0.42$.}
\label{fig:drop_tanner}
\end{figure}

Finally, we have already seen that the predictions of the resultant drop size obtained using the dynamic contact angle model are significantly more accurate than those obtained using the fixed contact angle model, and so it is not surprising that the results obtained from the numerical simulations are quite sensitive to the details of the dynamic contact angle model (parameters $\lambda$ and $n$) used (Fig.~\ref{fig:drop_tanner}). For a fixed value of $\lambda$, the drop size increases with $n$ at the same stretching speed and this behavior is more pronounced at higher stretching speeds. The size increase with $n$ can be explained from the structure of the model (Eq.~\ref{eq_tanner}) by the fact that the maximum angle variation, $(\theta - \theta_r)$, is less than one as the contact line recedes. For the same angle difference $(\theta - \theta_r)$, increasing $n$ causes the contact line speed $u_c$ to decrease, which, as we have seen above,  leads to a smaller resultant drop. As $n \rightarrow 0$ the contact line moves with less dependence on the contact angle, with which the drop size is expected to change slightly with the stretching speeds. For a fixed $n$, raising $\lambda$ causes drop sizes to decrease because the contact line speed $u_c$ is linearly proportional to $\lambda$. Numerical experiments with varying $\lambda$ for $n = 3$ shows that the drop size changes slowly with $\lambda$ as the contact line speed $u_c$ has a linear dependence on $\lambda$ but is dominated by the power term: $(\theta-\theta_r)^n$. In the limit of $\lambda \rightarrow 0$, the contact line becomes immobile, which is identical to the stretching bridge with a fixed contact line.

\section{Conclusion}
We have used experimental and theoretical methods to understand the role of volume and surface wettability in the breakup of a stretching liquid bridge with a moving contact line.  Unlike previous studies in which both contact lines are fixed, this system is strongly influenced by the details of the dynamic contact angle at the lower boundary, which controls the contact line motion and, through this, the point at which the bridge becomes unstable to pinchoff. The configuration has many practical applications associated with drop dispensing from a syringe (constant-pressure deposition), although in the current case drop deposition from a constant-volume liquid bridge was studied in order to simplify the problem.  Experimental measurements are reproduced with excellent accuracy using (i) a quasi-static analysis to predict the initial evolution of the bridge and the onset of contact line motion, (ii) a stability analysis to predict the onset of the rapid pinch-off of the column and (iii) a one-dimensional dynamical model, incorporating a variable contact line model, to predict the unsteady evolution of the bridge during the rapid pinchoff process.

For deposition from constant-volume bridges, a slight increase in the resultant drop size was observed as the bridge was stretched faster, an increase which can be attributed to the reduction in the time between the point at which the contact line starts to accelerate inwards and the point at which the liquid bridge breaks. Also drop size dependence on liquid volume was observed in experiment and can be explained (using the stability analysis) by the change in the critical radius as a function of the bridge volume. In addition, the effects of the equilibrium wetting angle on the critical radius were investigated within the framework of the the stability analysis and confirmed by both experiments and numerical calculations.

The combined effects of volume and dynamic break help to interpret the observed trend of the drop changing with the syringe speed in constant-pressure deposition, which shows a dramatic decrease in drop size at the beginning and down to a minimum the drop size slightly increases with the syringe speed. When the flow speed is still comparable to the syringe speed, the liquid volume determines the critical state of the liquid bridge and the role of the dynamic angle is insignificant. However, up to some point, the syringe speed is much higher than the flow speed and it can be assumed the volume change due to syringe speed is negligible. Similar to constant-volume cases, the drop size slightly increases with syringe speed. The competition between the effects of volume and dynamic contact angle leads to a minimum drop size at an optimized syringe speed.

We also quantitatively investigated the influence of the dynamic contact angle on drop dispensing by comparing the numerical results from calculations with conditions of fixed contact angle and dynamic variable contact angle. Although calculations with both conditions capture the essential features of the contact line motion, the numerical results from a dynamic contact angle model showed much better quantitative agreements with the experimental measurements. In the fixed contact angle calculation, the maximum contact line speed calculated is ten times higher than that observed in experiment, a discrepancy that leads to the prediction of smaller drop sizes that are observed in practice.  Although the numerical calculation with a dynamic contact angle model matches well with experiments, this agreement is achieved by adjusting the model parameters, and it should be admitted that there is no general criteria for choosing the model parameters which may change from case to case. Moreover, the 1D model can not accurately solve for the radial flow, especially the variation near the ends of the bridge. Accurate numerical prediction of the dispensing drop size requires solving the full (two-dimensional) governing equations with an accurate model of the dynamic contact line, which couples length scales from microscopic (dynamic contact line) to macroscopic(liquid bridge). Finally, Maragoni effect has not been taken into account within the numerical model and it possibly plays a role very near bridge pinch-off at which the rapid retraction of bridge surface may create a surface tension gradient along the bridge neck even without temperature gradient and surfactant.

\begin{acknowledgments}
We would like to thank David Gagnon and Melissa Loureiro for their assistance with the experiments.  The research was supported by the US National Science Foundation.
\end{acknowledgments}

\bibliographystyle{jfm}
\bibliography{drop}


\end{document}